\documentclass[aps,prl,showpacs,twocolumn]{revtex4-1}
	\usepackage{graphicx}
	\usepackage{bm}
	\usepackage{color,hyperref,graphicx,ulem,subfigure}
	\usepackage{amsmath, amssymb}
        \usepackage{epstopdf}

\begin{document}
	\title{Nonclassicality Phase-Space Functions: More Insight with Fewer Detectors}

	\author{A. Luis}
	\affiliation{Departamento de \'{O}ptica, Facultad de Ciencias F\'{\i}sicas, Universidad Complutense, 28040 Madrid, Spain }

	\author{J. Sperling}
	\affiliation{Arbeitsgruppe Theoretische Quantenoptik, Institut f\"{u}r Physik, Universit\"{a}t Rostock, D-18051 Rostock, Germany }
	\author{W. Vogel}
	\affiliation{Arbeitsgruppe Theoretische Quantenoptik, Institut f\"{u}r Physik, Universit\"{a}t Rostock, D-18051 Rostock, Germany }
	\date{\today}

\begin{abstract}
	Systems of on-off detectors are well established for measuring radiation fields in the regime of small photon numbers.
	We propose to combine these detector systems with unbalanced homodyning with a weak local oscillator.
	This approach yields phase-space functions, which represent the click counterpart of the $s$~parametrized quasiprobabilities of standard photoelectric detection theory.
	This introduced class of distributions can be directly sampled from the measured click-counting statistics.
	Therefore, our technique visualizes nonclassical effects without further data processing.
	Surprisingly, a small number of on-off diodes can yield more insight than perfect photon number resolution.
	Quantum signatures in the particle and wave domain of the quantized radiation field, as shown by photon number and squeezed states, respectively, will be uncovered in terms of negativities of the sampled phase-space functions.
	Application in the vast fields of quantum optics and quantum technology will benefit from our efficient nonclassicality characterization approach.
\end{abstract}

	\pacs{42.50.-p, 03.65.Wj, 42.50.Dv} 
                         
\maketitle

\paragraph{Introduction.--}
	Experimental methods of quantum state reconstruction opened new possibilities to visualize quantum states in phase space~\cite{SBRF93}; for more details see, e.g., the review articles~\cite{WVO99,LR09}.
	Since the description in phase-space is well established in classical physics, the corresponding representation in the quantum domain is very useful to determine the quantum effects of light and matter.
	In this context, the Glauber-Sudarshan representation of a general quantum state $\rho$,
	\begin{equation}
		\label{eq:P-repr}
		\rho = \int d^2\beta P(\beta) |\beta\rangle \langle\beta|,
	\end{equation}
	in terms of coherent states $|\beta\rangle$ with complex amplitude $\beta$, plays a central role~\cite{G63,S63}.
	The $P$~function of a coherent state agrees with the corresponding classical phase-space representation, for both coherent light and coherent degrees of freedom in matter systems.
	Any classical mixture of coherent states corresponds to a $P$~function having the properties of a classical probability density.

	A quantum state is said to be a nonclassical one whenever its $P$ function cannot be interpreted as a classical probability density~\cite{TG65,M86}.
	However, the function $P(\beta)$ can become strongly singular, so that it is not observable in general.
	In terms of distribution theory, nonclassicality requires that $P(\beta)$ has negativities which define the notion of a quasiprobability distribution.
	It has been theoretically shown~\cite{KV10} and demonstrated in experiments~\cite{KVBZ11,KVHS11,KVCBAP12} that a regularized version of the Glauber-Sudarshan $P$~function, can be recorded for any quantum state.
	Moreover, it uncovers all negativities included in the $P$~function.

	For the domain of weak radiation fields, when the photon statistics dominates the quantum phenomena, special measurement strategies have been developed.
	For example, the light is measured by subdividing it into equal intensity parts which are detected by avalanche photodiodes (APD) in the Geiger mode.
	Typically a number of coincidence ``clicks'' is recorded by the detector system, which indicates that at least this number of photons has been observed.
	The resulting click-counting statistics applies, for example, to detector arrays or time-bin multiplexing setups consisting of a given number of APDs; see, e.g.,~\cite{WDSBY04,ASSBW03,ZABGGBRP05,LFCPSREPW08,ABA10,ALCS10}.
	The click statistics of such devices differs significantly from the standard photodetection statistics~\cite{SVA12}.
	To identify quantum effects on this basis, such as sub-binomial light~\cite{SVA12s,Bartley}, proper nonclassicality criteria have been established~\cite{SVA13}. 
        The probability of single outcomes can also reveal the quantum nature of states and measurements~\cite{RL09}.

	In the present contribution we derive a method for the direct sampling of phase-space functions employing click-counting detector systems.
	Our technique yields a click counterpart of the $s$~parametrized quasiprobabilities of Cahill and Glauber~\cite{CG69}.
	Any negativity occurring in this click phase-space function for an even number of click detectors certifies the quantum character of the probed system.
	These nonclassicality conditions will be applied to squeezed states and photon number states, as typical representatives of the wave and particle nature of quantum light, respectively.
	Our method opens the possibility to detect phase-sensitive quantum effects in a regime that is usually used with on-off detectors to observe particle-like quantum effects.

\begin{figure}[h]
\begin{center}
	\includegraphics[width=7cm]{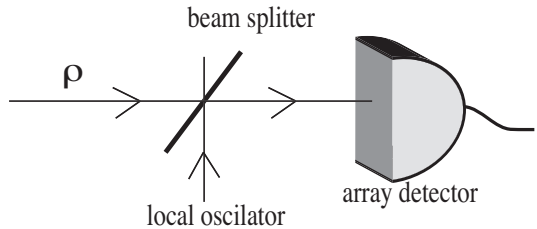}
	\caption{
		Unbalanced homodyne detection setup.
		A signal state $\rho$ is mixed on a beam splitter with a controlled local oscillator.
		A uniformly illuminated array of APDs registers the resulting click-counting statistics.
	}\label{fig:setup}
\end{center}
\end{figure}

\paragraph{Click-counting phase-space functions.--}
	The objective is to provide practical evidence of the negativity of the Glauber-Sudarshan $P$ function of nonclassical light by homodyne detection performed with on-off detector arrays.
	The rather simple experimental setup is shown in Fig.~\ref{fig:setup}.
	It is an unbalanced homodyne detection scheme, in which both the amplitude and phase of the local oscillator can be controlled.
	The array detector consists of $N$ on-off detectors which are equally illuminated.

	For analyzing measurements with click detector systems, it is important to apply the true click-counting statistics of these devices~\cite{SVA12}. 
	In the measurement setup under study, the signal quantum state $\rho$ is coherently displaced by the local oscillator.
	The resulting output state $\rho_\alpha$ is recorded by the array of $N$ on-off detectors with a quantum efficiency $\eta$.
	The latter efficiency is a product of the detection efficiency of a single APD and the transmittance of the beam splitter.
	The click statistics $c_k$, i.e., the probability that $k$ ($k = 0,1,\ldots N$) of the $N$ on-off detectors record a click, is given by
	\begin{equation}
		\label{eq:clickstat}
		c_k  \left ( \alpha ; \eta \right ) = \mathrm{tr} \left ( \rho_\alpha \Pi_{k,\eta} \right ) ,
	\end{equation}
	where 
	\begin{equation}
		\label{povm}
		\Pi_{k,\eta} = : \binom{N}{k}  e^{-\eta (N-k)n/N} \left ( 1 - e^{- \eta n /N} \right )^k :, 
	\end{equation}
	$n=a^\dagger a$ is the photon number operator of the signal state $\rho$, and ${:}\cdots{:}$ denotes normal ordering.

	Let us now combine the click-counting technique, which is phase insensitive, with the detection of quasiprobabilities in phase space.
	For this purpose we introduce a click counterpart, $P_{N} \left ( \alpha ; s \right)$, of the $s$~parametrized quasiprobabilities of Cahill and Glauber~\cite{CG69}.
	The index $N$ specifies the number of on-off detectors.
	It can be directly sampled by any click-counting device with quantum efficiency $\eta$ via
	\begin{equation}
		\label{Pp}
		P_{N} \left ( \alpha ; s \right ) = \frac{2}{\pi (1-s)} \sum_{k=0}^N \left [ \frac{\eta (1-s)-2}{\eta (1-s)} \right ]^k
		c_k  \left ( \alpha ; \eta \right ). 
	\end{equation}
	This definition is based on the method to sample the true $s$~parametrized distribution~\cite{WV96} by simply replacing therein the photoelectric counting statistics $p_k  \left ( \alpha ; \eta \right )$ with $k=0,\dots, \infty$ by the measured click statistics~\eqref{eq:clickstat} for a finite number of clicks, $k=0,\dots, N$.

	In the limit $N\to\infty$ the click statistics tends to  the photoelectric counting distribution, $\lim_{N \rightarrow \infty}  c_k   \left ( \alpha ; s \right ) = p_k  \left ( \alpha ; s \right ) $, cf.~\cite{SVA12}.
	As a consequence, in this limit $P_N$ tends to the true $s$-ordered quasiprobability distribution, $\lim_{N \rightarrow \infty} P_N \left ( \alpha ; s \right ) = P \left ( \alpha ; s \right )$~\cite{footnote1}, and hence it fully characterizes the quantum state under study.
	For finite $N$, after the replacement $p_k  \left ( \alpha ; \eta \right ) \longrightarrow c_k  \left ( \alpha ; \eta \right )$, Eq.~(\ref{Pp})  no longer provides the true $P \left ( \alpha ; s \right )$~function, but some regularized version of it -- as we will observe later on.
	Since the number of clicks is finite, there is no room for the pathological behavior in the form of derivatives of the Dirac delta distribution as it often occurs in $P \left ( \alpha ; s=1 \right )$ for nonclassical states.
	Thus we expect that $P_N \left ( \alpha ; s \right )$ reveals the nonclassical character of the field in an experimentally accessible form, without the obscurity provided by the singularities of the original $P$~function.
	
\paragraph{Visualizing nonclassicality.--}
	The approach of Refs.~\cite{KV10,KVBZ11,KVHS11} aims at visualizing the nonclassical effects covered in the often singular $P$~function in experiments.
	This requires the post-processing of the measured data by filtering techniques which are properly designed for this purpose.
	In the present approach we replace the measurement of the photoelectric counting distribution, which is hardly recorded by available measurement devices, by the click statistics of on-off detector systems.
	The statistical properties of the latter differ significantly from those of the photoelectric counting statistics.
	We will demonstrate that just this different statistics of the measured clicks makes the otherwise needed data post-processing superfluous.
	Hence the click-counting phase-space functions are suited to directly uncover the quantum effects of light. 

	Let us rewrite the click counting quasiprobability~(\ref{Pp}) for the purpose of visualizing quantum effects.
	Inserting Eqs.~(\ref{eq:clickstat}) and~(\ref{povm}), we obtain after some algebra
	\begin{align}
		P_N \left ( \alpha ; s \right )  =&  \frac{\eta}{\pi} \left[ \frac{2}{\eta (1-s)} \right]^{N+1} 
		\\\nonumber
		&\times \left \langle \!{:} \left [ e^{-\eta  n/N} + \frac{\eta(1-s)-2}{2}\right]^N {:}\! \right \rangle_{\rho_\alpha}.  
	\end{align}
	The notion $\langle \dots \rangle_{\rho_\alpha}$ is used for the quantum averaging with the displaced signal state $\rho_\alpha$, which is obtained by combination with the local oscillator, as shown in the detection scheme in Fig.~\ref{fig:setup}.
	It is worth mentioning that in this form $P_N$ is proportional to the generating function of the click statistics, which has been used to formulate nonclassicality criteria~\cite{SVA13}.

	Denoting by $P_\alpha(\beta)$ the $P$~representation~(\ref{eq:P-repr}) of the displaced state $\rho_\alpha$, the click-counting phase-space function can be written as
	\begin{align}
		P_N \left ( \alpha ; s \right )  {=}&   \frac{\eta}{\pi} \left[ \frac{2}{\eta (1-s)} \right]^{N+1} \\
		&\nonumber {\times} \int d^2 \beta P_\alpha ( \beta)\left [ e^{-\eta |\beta|^2/N} + \frac{\eta(1-s)-2}{2}\right]^N \!\!.
	\end{align}
	Let us further consider detector systems with an even number $N$ of on-off detectors, as it is usually the case in experiments of such a type.
	Thus, any negative value of $P_N \left ( \alpha ; s \right )$ can only arise from negativities of the function $P_\alpha(\beta)$.

	It is noteworthy that the coherent displacement of a quantum state is a purely classical operation. 
	As a consequence, any negativity of the displaced state $\rho_\alpha$ implies a corresponding negativity of the original signal state $\rho$ under study.
	For any classical state, the $P$ function and its displaced version, $P_\alpha$, are positive semi-definite.
	Hence, the same holds true for any click-counting quasiprobability $P_N \left ( \alpha ; s \right )$ with even $N$ and $s<1$, which can be directly sampled in experiments.
	The other way around, any verification of $P_N \left ( \alpha ; s \right )<0$, for some values of $s$, $\alpha$, and $N$ even, is a doubtless verification that the signal state $\rho$ is nonclassical and its $P$~function has negativities.
	This means that the state has no representation as a classical mixture of coherent states.
	Hence it must include quantum interferences between coherent states.

	The question may appear whether our approach identifies all the negativities of the Glauber-Sudarshan $P$~function or not.
	As discussed above, in the limit $N \to \infty$ the quasiprobability $P_N$ converges to $P$.
	Consequently, the information about the state increases with increasing number $N$ of APDs.
	As the sampling in our scheme is possible for any value of $s<1$ and finite $N$, in the domains $N\gg 1$ and $s\lessapprox1$, $P_N$ becomes asymptotically close to the Glauber Sudarshan $P$~function.

	Nevertheless, even if fewer detectors retrieve less information, they can gather it in a more efficient way regarding nonclassicality.
	We are going to see that even for small $N$ values the visualization of quantum effects becomes possible for quantum states whose original $s$~parametrized quasiprobabilities would fail to exhibit regular negativities.
	As typical examples we will study photon number and squeezed vacuum states.
	The former are representatives of the particle nature of quantum light, whereas the latter show phase-sensitive quantum effects reflecting the wave nature of the radiation field.

\paragraph{Photon number states.--}
	Let us start with the example of photon number states as the signal, $\rho=|n\rangle \langle n|$.
	Behind the beam splitter of the setup in Fig.~\ref{fig:setup}, the state -- being recorded by the array detector -- is the displaced number state,
	\begin{equation}
		\label{eq:displ-Fock}
		\rho_\alpha = D (-\alpha) |n\rangle \langle n| D^\dagger (-\alpha).
	\end{equation}
	The click statistics $c_k$ for this state is obtained from Eq.~(\ref{eq:clickstat}) together with~(\ref{povm}).
	Following Ref.~\cite{SVA14}, the positive-operator valued measure (POVM) of the click detector system can be given as
	\begin{align}\label{Eq:DsymClickPOVM}
		\Pi_{k,\eta} =\sum_{m=0}^\infty \mathcal D^{1-\eta,\eta}_{k,m} |m\rangle\langle m|.
	\end{align}
	The symbol ${\mathcal D}^{\tau,\sigma}_{k,m}$ is defined as
	\begin{align*}
		{\mathcal D}^{\tau,\sigma}_{k,m}=\binom{N}{k}\lim_{x\to0}\partial_x^m\left[{\rm e}^{\tau x}\left({\rm e}^{\frac{\sigma}{N}x}-1\right)^{k}\right],
	\end{align*}
	which can be efficiently computed through recursion relations, cf.~\cite{SVA14}.
	On this basis, the click-counting phase-space function $P_N(\alpha;s)$ is obtained from Eqs.~(\ref{eq:clickstat}), (\ref{Pp}), (\ref{eq:displ-Fock}), and~(\ref{Eq:DsymClickPOVM}).

	We will consider examples of photon number states with an odd photon number $n$.
	In this case, the negativities of the state occur in the origin of phase space, so that we can restrict our consideration to $P_N\equiv P_N \left( \alpha=0 ; s \right)$.
	To get some insight into the statistical significance of the accessible negativities, we consider a finite number $\nu$ of repetitions of the measurement.
	This yields the statistical error for the determination of $P_N \left( \alpha ; s \right)$ as
	\begin{align}
		\left[\Delta P_N \left( \alpha ; s \right)\right]^2 =&  \left[ \frac{2}{\pi (1-s)} \right]^2 \left(\sum_{k=0}^N  
		\left[ \frac{\eta (1-s)-2}{\eta (1-s)} \right]^{2k} \right. \nonumber\\
		&\left. \times  \frac{c_k  (\alpha ; \eta) \left[ 1- c_k  (\alpha ; \eta) \right] }{\nu} \right).
	\end{align}

\begin{figure}
\begin{center}
	\includegraphics[width=7cm]{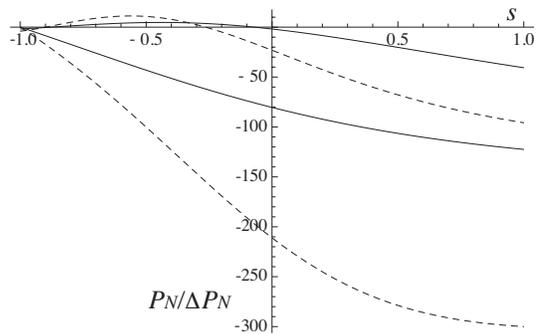}
	\caption{
	      Statistical significance $P_N / \Delta P_N$ as a function of $s$, for $n=1$, $N=4$ (lower curves), and $n=3$, $N=8$ (upper curves) for two quantum efficiencies $\eta=0.6$ (solid lines) and $\eta=0.9$ (dashed lines), for $\nu=10^4$ measurements.
	}\label{fig2}
\end{center}
\end{figure}

	In Fig.~\ref{fig2} we show the statistical significance $P_N / \Delta P_N$ as a function of $s$ for a 
	single-photon state, $n=1$, for $N=4$ APDs, with  $\nu = 10^4$ repetitions and for two quantum efficiencies.
	We also consider the situation for $n=3$ and $N=8$ on-off detectors.
	It is seen that the significances of the negativities increase with increasing the values of $s$ and the quantum efficiency $\eta$.
	This is in full agreement with standard expectations.

\paragraph{Squeezed vacuum states.--}
	Let us consider now the situation for the signal being in a squeezed vacuum field
	\begin{equation}
		\label{sv}
		| \xi \rangle = e^{r ( a^2 - a^{\dagger 2})/2} | 0 \rangle ,
	\end{equation}
	where $| 0 \rangle$ is the vacuum state and $r$ the squeezing parameter.
	The displaced light field is therefore a squeezed coherent state,
	\begin{equation}
		\label{eq:displ-sq}
		\rho_\alpha = D (-\alpha) |\xi\rangle \langle \xi| D^\dagger (-\alpha).
	\end{equation}
	The further analysis of the nonclassicality click phase-space function follows the same calculus as for the photon number states.

\begin{figure}
\begin{center}
	\includegraphics[width=7cm]{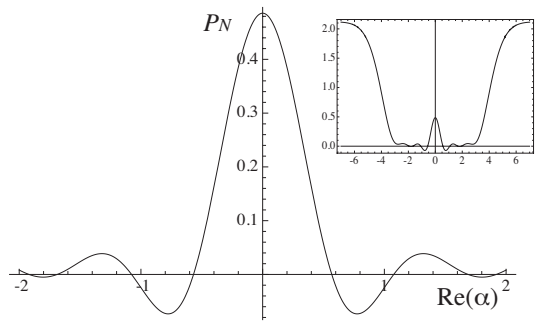}
	\caption{
		 $P_N (\alpha; s=0)$ for squeezed vacuum with $r=1$ as a function of $\mathrm{Re} (\alpha)$, for $\mathrm{Im} ( \alpha) =0$, $\eta =0.9$, $N=6$.
		 The inset shows the same function but for a larger range of $\mathrm{Re} (\alpha)$  values.
	}\label{fig:sq1}
\end{center}
\end{figure}

\begin{figure}
\begin{center}
	\includegraphics[width=7cm]{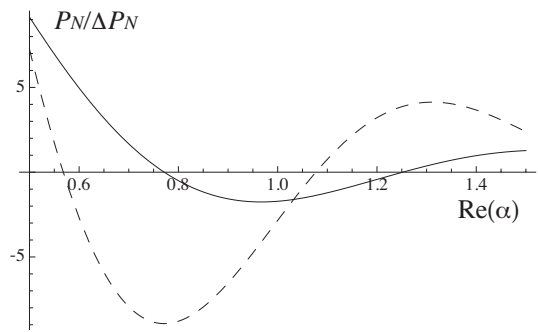}
	\caption{
		Statistical significance $P_N  / \Delta P_N$ after $\nu = 10^4$ measurements for the same parameters as used in Fig~\ref{fig:sq1} (dashed line) and for $s=-0.25$, $\eta=0.6$ (solid line).
	}\label{fig:sq2}
\end{center}
\end{figure}

	In Fig.~\ref{fig:sq1} we show $P_N (\alpha; s)$ for $r=1$, $N=6$, $\eta = 0.9$, and $s=0$, as a function of $\mathrm{Re} (\alpha)$ for $\mathrm{Im} (\alpha) =0$.
	It clearly shows the existence of regions of negativities of  $P_N (\alpha; s)$. 
	Restricting $\mathrm{Re} (\alpha)$ to the interval $[-2,2]$, the graph resembles Fig.~\ref{fig:sq2} in Ref.~\cite{KVHS11} for a regularized squeezed vacuum with a similar squeezing parameter.
	Interestingly, this similarity holds for the choice of $s=0$, which yields the Wigner function in the limit $N\to \infty$.
	It is noteworthy that the Wigner function of squeezed light, in general, has the properties of a classical Gaussian probability density.
	Hence, it is just the reduction of the number $N$ of on-off detectors, in this example to $N=6$, which uncovers the quantum effects in terms of negative values of $P_N (\alpha; s=0)$.
	Note that even a simple experimental setup with only two on-off detectors can be used to visualize by our method the quantum nature of a squeezed state.

	The inset in  Fig.~\ref{fig:sq1} shows the same function for $\mathrm{Re} (\alpha)$ in a larger domain.
	We clearly observe a saturation of the click phase-space function for increasing local oscillator amplitude $\alpha$.
	This behavior is actually not surprising.
	For sufficiently high intensities, all the on-off detectors click coincidentally and, therefore, saturate the sampled click-counting phase-space function.
	This means that the latter only formally resembles the structure of a quasiprobability for sufficiently small arguments $\alpha$ (this is for weak local oscillators in the homodyne array in Fig.~\ref{fig:setup}), because it cannot be normalized.
	This is, however, no fundamental constraint for the usefulness of our method for visualizing the quantum nature of light.

	The statistical significance of the negativities is shown in Fig.~\ref{fig:sq2}.
	The best negativity is found for  $\mathrm{Re} (\alpha)\approx0.8$.
	For the rather small number of $\nu = 10^4$ repetitions of the measurement, the statistical significance at this optimum point is $P_N/ \Delta P_N \simeq - 8.8$.
	We even can identify some negativities for the parameter $s=-0.25$ and the reduced quantum efficiency of $\eta=0.6$.
	The significance can be easily increased by increasing the number $\nu$ of data points.
	This result is quite promising, as for $s<0$ the limit $N\to \infty$ yields a quasiprobability which is more noisy than the Wigner function.
	Moreover, the reduced efficiency also increases the noise level.
	For a large number of APDs such distributions will not be useful to visualize quantum phenomena; however, this is possible for small values of $N$.

\paragraph{Conclusions.--}
	We have introduced the concept of phase-space functions -- the click counterpart to the prominent $s$~parametrized phase-space distributions of Cahill and Glauber -- for unbalanced homodyne measurements with arrays of on-off detectors.
	Any negativity of this function is an authentic indicator of the quantum character of the measured field.
	Because of the finite number of detectors, the sampling is also possible in the parameter range $0<s<1$, which would be impossible even if a perfect, true photon counter would exist.
	Our method is an easily accessible approach to visualize quantum effects of light in terms of negativities within phase-space functions, without the need of post-processing of the data as in the case of nonclassicality filtering.
	In the limit of an infinite number of on-off detectors, this allows us to identify quantum signatures of any quantum state.
	However, we demonstrated that seeking this limit might not be advantageous.
	A small number of diodes can demonstrate nonclassicality, which cannot be directly visualized by negativities of any $s$~parametrized phase-space quasiprobability.
	Our proposed characterization method of quantum phenomena may have a plethora of applications in rapidly growing research fields, such as quantum optics and quantum technology.

\paragraph*{Acknowledgments.--}
	This work was supported by the Deutsche Forschungsgemeinschaft through SFB 652.
	A. L. acknowledges support from project FIS2012-35583 of the Spanish Ministerio de Econom\'{\i}a y Competitividad.

\end{document}